TYPE: Full Paper/Research Progress

# LitStoryTeller: An Interactive System for Visual Exploration of Scientific Papers Leveraging Named entities and Comparative Sentences

Qing Ping[1]   Chaomei Chen[2]

[1] qp27@drexel.edu
Drexel University, Philadelphia (America)

[2] cc345@drexel.edu
Drexel University, Philadelphia (America)

**Abstract**
The present study proposes *LitStoryTeller*, an interactive system for visually exploring the semantic structure of a scientific article. We demonstrate how *LitStoryTeller* could be used to answer some of the most fundamental research questions, such as how a new method was built on top of existing methods, based on what theoretical proof and experimental evidences. More importantly, *LitStoryTeller* can assist users to understand the full and interesting story a scientific paper, with a concise outline and important details. The proposed system borrows a metaphor from screen play, and visualizes the storyline of a scientific paper by arranging its characters (scientific concepts or terminologies) and scenes (paragraphs/sentences) into a progressive and interactive storyline. Such storylines help to preserve the semantic structure and logical thinking process of a scientific paper. Semantic structures, such as scientific concepts and comparative sentences, are extracted using existing named entity recognition APIs and supervised classifiers, from a scientific paper automatically. Two supplementary views, *ranked entity frequency view* and *entity co-occurrence network view*, are provided to help users identify the "main plot" of such scientific storylines. When collective documents are ready, *LitStoryTeller* also provides a *temporal entity evolution view* and *entity community view* for collection digestion.

**Conference Topic**
Mapping and visualization; Knowledge discovery and data mining; Methods and techniques.

**Introduction**
With the sheer volume of scientific publications every year, it becomes a double-challenge for researchers to not only comprehend a collection of research articles as a whole, but also to grasp effectively important pieces of information scattered everywhere in each single article.
As a solution to this double-challenge, researchers from multiple areas have contributed insights. In the domain of scientific mapping, some existing work have proposed applications to digest a collection of research papers on collection-level, such as CiteSpace (Chen, 2006), Action Science Explorer (Dunne, Shneiderman, Gove, Klavans, & Dorr, 2012), VOSViewer (Van Eck & Waltman, 2010). In broader scope of digital humanity, several applications have been developed to digest a text corpus on topic-level, such as VarifocalReader (Koch, John, Wörner, Müller, & Ertl, 2014), Serendip (Alexander, Kohlmann, Valenza, Witmore, & Gleicher, 2014), on sentence-level, such as PICTOR (Schneider et al., 2010), and on word-level, such as POSvis (Vuillemot, Clement, Plaisant, & Kumar, 2009) and Wordle (Viegas, Wattenberg, & Feinberg, 2009).
Existing work mentioned above are insufficient to solve the double-challenge for scientific paper digestion. First, scientific mapping applications focus on extracting collection-level patterns as a whole, and are not suitable for individual document analysis. Second, applications in digital humanity, though on multiple-levels, are not tailored for scientific paper digestion. Most existing work in this area are designed for special text corpus, such as poem, play, news, Bible, and so on, but very few if not none are tailored for scientific papers. Third, even for those applications that are not confined to one type of text, the toolkit developed for detailed investigation is still simplified.

To bridge this gap, we present LitStoryTeller for better support of scientific paper digestion. On document-level, LitStoryTeller automatically extracts scientific concepts (or entities exchangeably in this paper) from full-text and visualizes entities and their co-occurrence and comparative relations in storylines. Here we use the visual metaphor of "storyline" in a play, where entities are considered as "characters", and paragraph/sentence are seen as "scenes" where "characters" get on stage. With this visual metaphor, we are able to preserve the logical plot of a scientific paper. Moreover, this storyline is synchronized with a text viewer, so that user could navigate through the full-text using the "characters" and "scenes" in the storyline as anchors. Supplementary views are also provided to help users to get focused on the main plot of the storylines. On collection-level, LitStoryTeller visualizes all entities in a collection with two different views, i.e. entity community view and temporal entity evolution view.

To our best knowledge, this paper is among the first work that is designed to support document-level exploration using a storyline visual metaphor and leveraging a variety of techniques such as entity extraction and comparative sentence classification methods. The main contributions of the present work are as follows:

1. We develop a framework for document-level exploration of scientific papers, using a "storyline" visual metaphor that preserves the logical thinking plot of a scientific paper;

2. We develop modules for named entity recognition and comparative sentence classification that could run in real-time to support semantic-level exploration of a scientific paper;

3. We also support collection-level exploration of scientific papers, using techniques for community detection algorithm and temporal network visualization.

The rest of the paper is organized as follows. *Related work* discusses existing work; *System design* presents the design of the proposed system; *Case study* demonstrates the utility of the system through a case study; *Conclusion* draws conclusion of the present study.

**Related Work**

*Single-document visualization*

In digital humanity, previous work has been done to facilitate users' exploration of a single document. Depending on the granularity of the visualization, applications can be divided into the following categories.

Topic-level document visualization

To facilitate exploration of a document, some applications focus on finding the latent topics of a document first, and use topics as an intermediary between words and full document for visualization. Varifocal-Reader (Koch et al., 2014) uses text segmentation method to segment full text into topical segments, and annotates entities such as person and location. Serendip (Alexander et al., 2014) uses statistical topic models as a bridge between words, topics and documents, and visualize the three elements in matrices. The advantage of this approach is that it helps to capture the topical structure of a document for easier digestion. However, it might also suffer from loss of finer-level details, such as detailed information in sentences and entities.

Sentence-level document visualization

Other applications focus on organizing and visualizing a document on sentence-level. One application chooses not to display all sentences plainly, but rather to display sentences using a fish-eye view so that salient passages will be highlighted as focal, and the rest will be put as receded (Correll, Witmore, & Gleicher, 2011). Another application extracts quote sentences from news narratives and could support searching of quotes by speakers (Schneider et al., 2010). The strength of this approach is that finer-level details (sentences) could be organized and shown to users. On the other hand, the weakness is that sentences alone cannot bear semantic meanings, unlike a topic or an entity.

### Word-level document visualization

There are multiple applications on word-level document visualization. One application finds frequent word usage patterns and highlights them in full-texts (Don et al., 2007). Another application supports to visualize all neighbouring words of a given word query in a word cloud view (Vuillemot et al., 2009). One work visualizes the word frequency distributions over the narrative scope (Clement, Plaisant, & Vuillemot, 2009). Another work proposes to visualize words in a document as word cloud, known as "Wordle" (Viegas et al., 2009). One work, specially tailored for play script, visualize characters-scenes as a matrix, with character-on-stage-scene as highlights (Wilhelm, Burghardt, & Wolff, 2013). There are also some works on phonetic-levels, often tailored for poem analysis (Abdul‐Rahman et al., 2013; McCurdy, Lein, Coles, & Meyer, 2016).

The advantage of visualization on word-level is that it reserves the finest-level of details. However, most of these works do not reserve the relationships between words, or entities.

Besides individual weaknesses, work mentioned above are not tailored scientific paper exploration. First, most of the work above are confined in specific corpus such as poem, play, news, Bible, and so on. Second, few of the existing work on word-level focus on word-word relationships. Third, existing work lacks support for semantic information extraction, such as comparative sentence classification.

### Scientific fields evolution visualization

Our present work is also analogous to the research of visualization of scientific field evolution on an abstract level. Research in this direction usually constructs a network of concepts or keyphrases by various proximity metrics based on co-word analysis, and then clusters the concepts into scientific fields. Then temporal patterns are investigated, such as the emergence and recombination of each scientific field in the network over time (Chavalarias & Cointet, 2013), and interactions between academic push and technological pull for theories (Callon, Courtial, & Laville, 1991). Here each scientific field is considered a "unit" in the storyline of scientific evolution, whereas in our present study, the basic "unit" of the storyline is a concept within a single scientific paper. Also, a link in scientific evolution represents the high-level connection between two scientific fields, whereas in our study, a link represents the sentence/paragraph-level co-occurrence of two concepts.

### Argumentation visualization

Our work may also be overlapped with research in argumentation visualization. Research in this area attempts to visualize the structures of argumentations, usually in an interactive collaborative learning environment, to support decision making (Kirschner, Buckingham-Shum, & Carr, 2012). Our work instead attempts to visualize the structures of concepts in a scientific paper via automatic natural language processing of the full text and interactive visualizations.

### Storyline visualization

There are two classical works in storyline visualization. One work proposes to visualize the storyline of a screen play by arranging characters and scenes over time (Tanahashi & Ma, 2012). More specifically, each character is represented by a curved line, and each scene is emphasized by bundling all character lines of this scene closer. Another work improves on the previous one by optimizing several objective functions to make the storyline more compact and visually-pleasing (Liu, Wu, Wei, Liu, & Liu, 2013). The present study utilizes a similar design of storyline as the bare-bone template(Elvery, 2017),which use curved line to represent a character, and a rectangle with lines passing through it to represent scenes. Character lines are arranged by first grouping characters using community detection algorithm, and then place each group as far apart as appropriate.

*Named entity recognition*

Majority of existing work on named entity recognition are supervised methods. Models used include Hidden Markov Models (HMM)(Bikel, Miller, Schwartz, & Weischedel, 1997), Maximum Entropy Models (ME) (Borthwick, 1999) and Conditional Random Fields (CRF)(McCallum & Li, 2003). Supervised named entity recognition usually have superior performances. However, these methods usually require a large amount of human-labeled data in a specific domain. In the present paper, we take advantage of the Microsoft Entity Linking API (https://www.microsoft.com/cognitive-services/en-us/entity-linking-intelligence-service), which not only recognizes named entities under a wide range of topics based on Wikipedia coverage, but also links entities of variant forms together.

*Comparative sentence classification*

Bing Liu has worked on the topic of comparative sentence extraction in a series of papers. In one paper, he proposes to use manual keyword list and frequent sequence mining, together with supervised classifier to classify sentences (Jindal & Liu, 2006a). In another work, he further proposes to not only classify sentences into comparatives/non-comparatives, but also extract the subjects as well as comparative relations from comparative sentences (Jindal & Liu, 2006b). In the present paper, we implemented the full pipeline for comparative sentence classification as in (Jindal & Liu, 2006a).

**System design**

In this section, we first provide an overview of the workflow of the proposed system, and then describe each module in detail.

*Overview of system workflow*

The overview of system workflow is depicted as in Figure 1. The workflow starts with a paper-uploading page. The uploaded full-text is tokenized, POS tagged, stemmed and converted to a feature vector. Then sentence feature vectors are sent to the comparative-sentence-classifier (① in Figure 1) to generate comparative/non-comparative labels. In the meanwhile, POS tagged and stemmed sentences are sent to the named-entity-recognition module (② in Figure 1). The recognized entities are sent to the single-document-storytelling module (③ in Figure 1). This module utilizes the labels generated in ① and entity-sentence-paragraph alignments generated in ② to create a storyline visualization, together with two supplementary views. Finally, when multiple papers have been uploaded into the system, the collective-documents exploration module (④ in Figure 1) can retrieve all entity co-occurrences stored in the local repository and visualize the entire document collection with two views. We will discuss each module below in detail.

*Comparative-sentence-classification module*

A comparative sentence expresses an ordering relation between two sets of entities with respect to some common features (Jindal & Liu, 2006b). Existing packages such as nltk.corpus.reader.comparative_sents (Pantone, 2017) cannot be used in the present study because it can only work on specific corpus with labeled comparative/non-comparative classes, entity names and relations. Instead, in the current system, we built a comparative-sentence-classification pipeline by implementing and modifying algorithms proposed in paper (Jindal & Liu, 2006a).

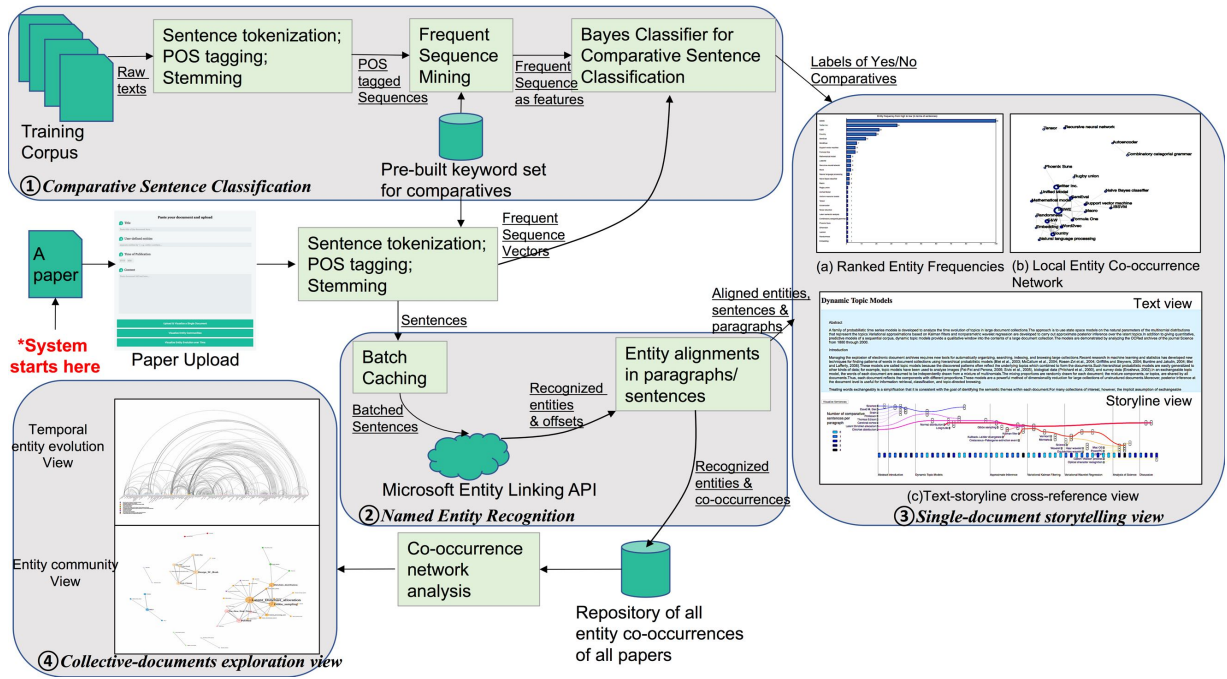

**Figure 1. Overview of LitStoryTeller system workflow**

## Full-text pre-processing

The objective of pre-processing is to convert the raw full-texts into POS tagged and stemmed sentences with corresponding indexing in paragraphs. After removing irregular characters and segmenting full-text into paragraphs by line breaks, we segment each paragraph to sentences with the NLTK PunktSentenceTokenizer. In the meanwhile, we record sentence offsets, namely starting and ending positions in paragraphs.

**Feature extraction**. The feature extraction step aims to pre-mine all frequent sequence patterns emerging from the training corpus. A frequent sequence pattern is a $(t_1, t_2, t_3, \ldots, t_k\_keyword, \ldots, t_{n-2}, t_{n-1}, t_n)$ sequence pattern with sufficient confidence and support in the corpus, where $t_i$ is POS tag of a surrounding word to a keyword. For example, assume the sentence "*X outperforms Y*" is prevalent in a corpus, then the corresponding frequent sequence pattern will be <*{NN}{VBZ, outperform}{NN}*>, where *outperform* is the keyword. More specifically, we perform the following steps to extract frequent sequence patterns:

(1) **Construction of keyword-list**. In the original work by (Jindal & Liu, 2006a), to identify comparative sentences, three categories of keywords are proposed, namely adjectival/adverbial comparatives, single-verb keywords, and phrase-keywords, 83 keywords in total. In our study, we added four keywords: "fail", "gain", "over" and "contrast".

(2) **Extraction of candidate sequences**. For each keyword matched in a sentence, we extract a sequence with certain window size for it as a candidate sequence. For example, for the following sentence, where *outperform* is in our keyword-list:

"*The concatenated features A **outperform** the original feature set of B.*"

The corresponding candidate sequence will be (window size = 3):

[('JJ'), ('VBZ'), ('DT'), ('outperform', 'NN'), ('DT'), ('JJ'), ('NN')]

(3) **Frequent sequence pattern mining**. We adopt the PrefixSpan algorithm to mine frequent sequence patterns from all candidate sequences generated in the last step. The PrefixSpan algorithm utilizes projection of search space into prefix sequences to reduce the number of candidate subsequence generations (Han et al., 2001). In our study, we implemented the PrefixsPan algorithm, with the minimum support $\tau$ for frequent sequence set to be 0.1, and the minimum confidence set to be 0.6.

**Classifier training**. After getting all the frequent sequence patterns, we consider them as features of a sentence, and train a Bayes Classifier based on given labels (Jindal & Liu, 2006a). If a sentence satisfies one frequent sequence pattern, the corresponding feature will have value 1, otherwise 0. We manually labelled 286 sentences from research papers, and feed the training corpus into a classifier. Here we report the results of Bayes classifier, with higher precision than those of SVM and Logistic Regression classifier. The accuracy of 5-fold cross-validation is $(0.84 \pm 0.02)$ for the Bayes classifier.

**New sentence prediction**. In the prediction-phase, each sentence is first POS tagged and stemmed. Then the sentence is converted to a feature vector indicating what frequent sequence patterns this sentence satisfies. Last, the feature vector is fed into the trained Bayes classifier to generate a prediction, namely comparative or non-comparative.

*Named-entity-recognition module*

We refer entities as important scientific concepts or terminologies discussed in a research paper. To recognize such entities in a researcher paper on-the-fly with high recall is a challenging task. In the current system, we take advantage of the Microsoft Entity-Linking API. This API not only recognizes a wide range of scientific concepts recorded in Wikipedia, but also automatically performs entity disambiguation.

One down-side of using the Microsoft Entity Linking API is that there is a limit for the service per day. Therefore, we propose a batch mechanism to minimize the number of calls as few as possible for each paper. We batch sentences into blocks before sending them using API. The block size is set to be 10000 characters. Then we use the returned results from the API to map the entity offsets in the block back into its offsets in sentences and paragraphs.

*Single-Document Story Telling*

This part of the system focuses on supporting users to grasp the important pieces of detail information in a research paper through storyline visualization. First, to help users started in the storyline, we provide two ways to determine which entities to read first ( Figure 1(a) and (b)). Following these leads, user could read the storylines of these entities through the "Text-storyline cross-reference" view ( Figure 1(c)) iteratively and spot important scientific conclusions with the help of "comparative sentence indicators".

Ranked entity frequency view

This view provides an intuitive way is to determine which entities to explore first: to look at how many times an entity has been mentioned in a document as in ( Figure 1(a)). Users could start their exploration focusing on these top-ranked entities.

Local entity co-occurrence network view

We could also consider the importance and interestingness of an entity to be its authority over other entities as in view ( Figure 1(b)). We use the force-layout(Bostock, 2017) to draw the co-occurrence network of entities. The size of the entity indicates its frequency in a document; while the width of link indicates how frequently two entities co-occur in one sentence.

Text-storyline cross-reference" view

After determining which entities are of high priority to explore first, users could use the "text-storyline cross-reference" view for "close-reading". More specifically, we design two views within the "text-storyline cross-reference" view, namely, the "Storyline view" and "Text view" ( Figure 1(c)). The collaboration between the two views will help users to navigate through the full text following accumulative clues in the storyline viewer.

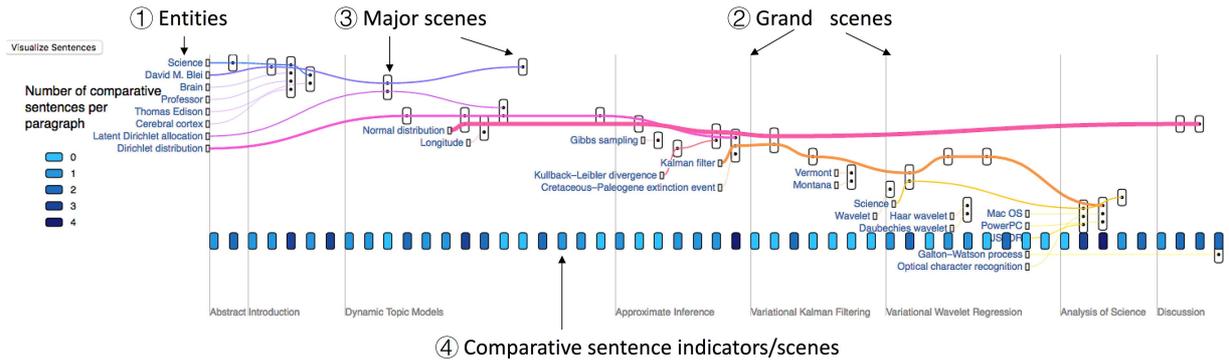

(a) Initial storyline of a research paper

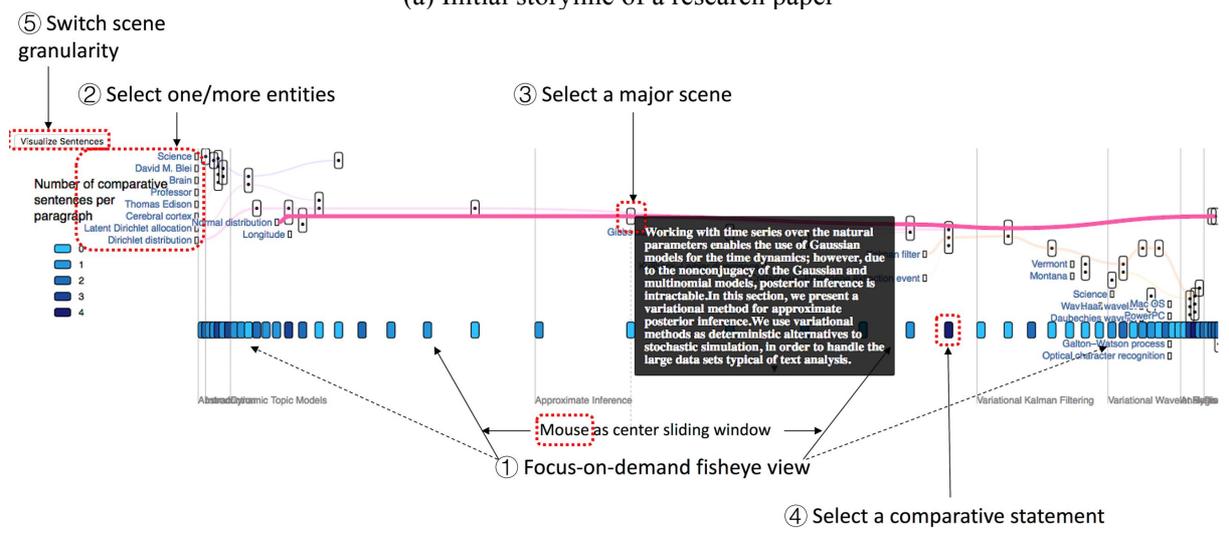

(b) Re-drawn storyline with focus-on-demand Cartesian fisheye view after mouse move

**Figure 2. Storyline of a research paper**

*Storyline View*

We borrow the "storyline" metaphor from theatre play for our visualization of a research paper. As in traditional "storyline" of a novel or a play, there are characters and scenes, and there are beginning, development, turning point, climax and conclusion in the plot.

A research paper shares similarity with the formality of a play. Essentially, a research idea is demonstrated through developments of concepts throughout its abstract, introduction, related work, methodology, experiment, discussion, conclusion and references. These sections can be regarded as grand scenes at a coarse-level. Further, each paragraph and each sentence can be considered as major scenes at fine-level. In our study, the scientific concepts are referred as entities, which can be extracted automatically from each sentences, paragraphs, and sections.

As in
**Figure 2**(a), the storyline of a research paper comprises of entities (characters) and sections/paragraphs/sentences (scenes). The storyline should be read from left to right, as its development in the research paper. More specifically, the glyphs in the storyline represents different elements as follows:

①**Entities**. Each entity, in a unique colour, is a scientific concept, represented by a curved line called lifeline, from left to right, indicating the period that the entity is first/last mentioned. The width of the curve line is in proportion to the frequency of an entity.

②**Grand scenes**. Sections, such as abstract, introduction, or conclusion, are considered as grand scenes. These grand scenes are visualized with vertical lines separating the storyline into sections.

③**Major scenes**. Each major scene is a paragraph or sentence with at least one recognized entity in it. It is represented by a transparent rectangle with soft corners. Each entity lifeline may pass through one or more major scenes, indicating occurrences in these major scenes. Multiple entity lifelines passing through the same major scene indicate their co-occurrences.

④**Comparative sentence indicators/each scene**. At the bottom of the storyline, we visualize every scene with a rectangle at equal steps. The shade of the rectangle represents our confidence whether this scene contains "comparative statements".

As in **Figure 2**(b), the storyline visualization is the main vehicle for users to perform various operations and navigate through the full-text document. We have enabled the following interactive operations on the storyline viewer:

①**Focus-on-demand fisheye view**. To avoid too long and sparse a storyline, a focus-on-demand zoom-in/zoom-out mechanism is desirable. We support this goal by implementing a fish-eye effect on the storyline (**Figure 2**(b)-①). This enables users to expand any part of a storyline to see details at focus, and shrink the rest of the storyline as background.

②**Select one or more entities**. The label as the starting point of each entity is selectable (**Figure 2**(b)-②). When clicked, the corresponding life-line of entity will be highlighted.

③**Select a major scene**. Each major scene, represented with rectangle, is selectable (**Figure 2**(b)-③). When clicking the major scene, the "Text view" will synchronically jump into and highlight the corresponding section of the scene ( Figure 1-①).

④**Select a comparative statement**. Each scene is displayed at the bottom with equal steps (**Figure 2**(b)-④), whose shade indicates how likely the scene contains a comparative statement. These glyphs are also selectable and synchronized with the text viewer.

⑤**Switch scene granularity**. The granularity of the scene can be switched from "paragraph" to "sentence" by clicking on the "visualize sentences" button (**Figure 2**(b)- ⑤).

*Text view*

We enable the text view to have synchronized responsive behaviour in accord with operations in the storyline view. That is, when corresponding operations be performed in the storyline viewer, such as selecting an entity, selecting a scene (paragraph/sentence), the corresponding content within the text view will be highlighted and brought to the focus area.

*Overview of a collection*

We provide two types of visualizations: temporal entity evolution view and entity community view.

Temporal entity evolution view

The temporal entity evolution view is designed to visualize how entities in an entire document collection evolve over time. Some entities appear very early in the collection, marked by the publication date of its corresponding paper, while others emerge at much later stage, with strong connections to the earlier ones. These connections can also be seen as mapping between research fronts (novel entities) and intellectual base (old entities) (Chen, 2006). We take co-occurrences between old and novel entities on sentence-level as the connections, and visualize these connections chronologically to show entity evolution.

As in Figure 1-④, we use an arc-diagram layout to visualize entity evolution over time. In our arc diagram, each circle represents an entity with label below it. Its colour indicates the corresponding paper where this entity first appeared in. Each arc connecting two circles represents co-occurrences of two entities in one or more sentences, whose thickness indicates the frequency of the co-occurrences. The entities are arranged from older to new horizontally based on their corresponding papers' publication time where they first appeared in.

Entity community view

The entity community view is designed to visualize the communities within the co-occurrence network of entities across the collection. In entity communities, some entities appear as outliers, while others connect multiple other entities to form a cohesive community, or occupied important "positions" in the network such as connecting two major communities as pivotal points (Chen, 2004). These entities are worth further analysis.

As in Figure 1-④, we implement a force-layout graph (Bostock, 2017) to visualize the entire entity co-occurrence network. The force-layout attempts to minimize the number of crossings of edges in a network by optimizing energy functions(Kobourov, 2012). In our visualization, each circle represents an entity, and its size indicates its frequency in entire document collection. Each link represents co-occurrences between two entities, whose width indicates how often the two entities co-occurred together. We use the Louvain method for community detection which optimizes global modularity through updating local communities(Blondel, Guillaume, Lambiotte, & Lefebvre, 2008). Entities belonging to one community are coloured with a unique colour.

**Case study**

We report a case study to demonstrate the purpose and functions of the current system design.

*Use Scenario*

We present the use case of a typical user named Alice. Alice is a graduate student, who has previous experience in research, and has some exposure to interactive visual analytic systems. Alice is surveying on research papers related to the topic of "Latent Dirichlet Allocation" (LDA), a probabilistic graphical model (Blei, Ng, & Jordan, 2003). Alice has already obtained a collection of full-text research papers on this topic (13 papers). Alice realizes that there are multiple models before and after the proposal of LDA, such as LSI, pLSI, hLDA and HDP. Alice is especially interested in how each model was built on top of one or more previous models, by improving the weakness of the previous ones on various metrics. Alice is also curious about what accompanying algorithms are frequently used together with these topic models, such as inference methods. These research questions are summarized as follows:

(1) *How is each novel model built or different from old models? Taking the example of LSI and pLSI, how does the new model (pLSI) is superior to the old model (LSI)?*
(2) *What mathematical formula and inference algorithms are usually used together with each topic model?*
(3) *Overall, what concepts (topic models, mathematical formula, and inference algorithms) are discussed the most in this document collection? How do these concepts evolve over time?*

*Single-Document Storytelling*

To answer research question (1) and (2), Alice selects the paper of probabilistic Latent Semantic Indexing (pLSI) to see how pLSI is built on top of Latent Semantic Indexing (LSI). Alice uses the single-document storytelling view for this task.

Due to space limitation, we skip the step of finding important entities with our supplementary views, and assume Alice already identifies the most important entities. Recall that question (1) is "How is each novel model built or different from old models?" To answer this question, Alice decides to read the storylines of the two main characters: pLSI and LSI at paragraph-level. Alice observes that the two highlighted lifelines have several crossings (co-occurrences in a paragraph), as in Figure 3. Alice believes that these crossings are crucial to answer question (1). Interestingly, all crossings yield important information about the relationships between pLSA and LSI, as can be read from the tooltip of the crossings in the figure. These tooltips clearly stated how pLSA is built on top of LSI theoretically and mathematically.

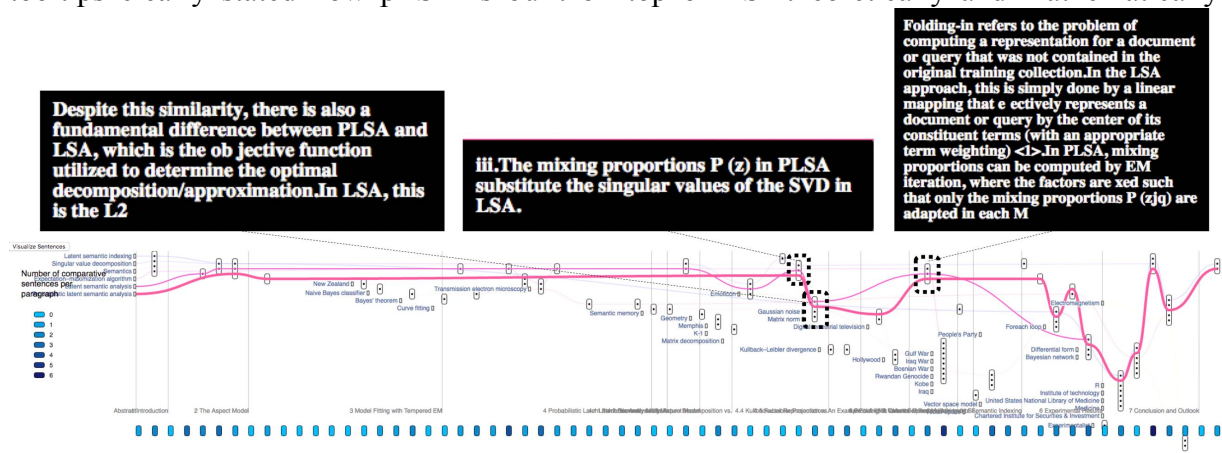

**Figure 3.Storyline of "Probabilistic Latent Semantic Indexing" at paragraph-level.**

Now Alice wants to go deeper into details, so she switches to read the storyline at sentence-level, as in Figure 4. This time, Alice notices multiple crossings at the "6. Experimental Results" section. Alice decides to read the sentences in these crossings. Interestingly, all crossings yield "conclusive" information about the relationships between pLSI and LSI, as can be read from the tooltips. Also, most of the crossings belong to the scenes with darker shade at bottom, indicating that pLSI and LSI are frequently compared in the section of "Experimental Results". At this point, Alice is very confident about knowing how pLSA is different and most likely an improvement from LSI, with both theoretical foundations and experimental results.

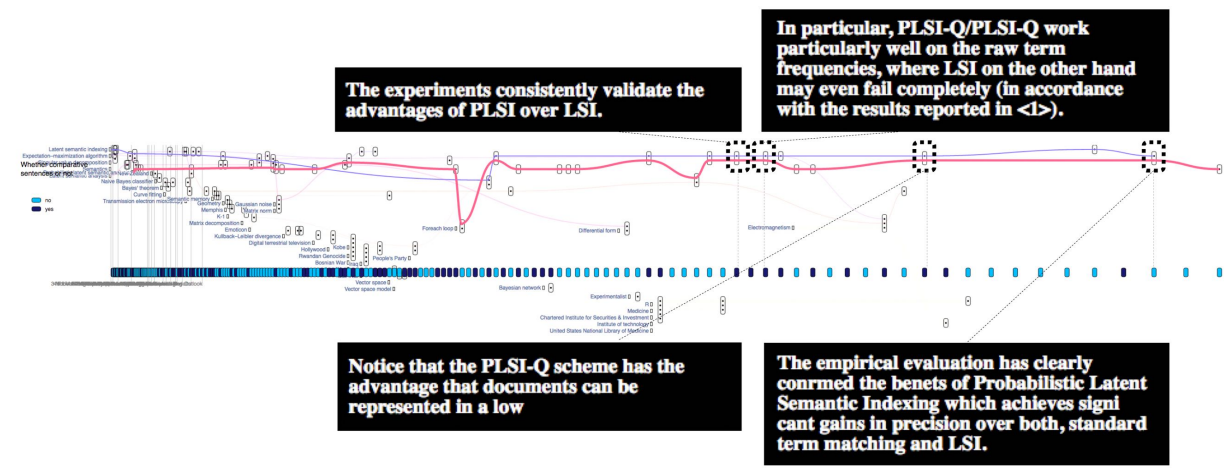

**Figure 4.Storyline of "Probabilistic Latent Semantic Indexing" at sentence-level.**

To answer research question (2), Alice examines crossings in Figure 4 again. Alice finds out that LSI is mostly associated with "Singular Value Decomposition", and pLSI with "Expectation-Maximization Algorithm". Indeed, to solve LSI, the most common used method is "Singular Value Decomposition", while for pLSI, the algorithm for inference is "Expectation-Maximization Algorithm".

*Collective-document exploration view*

To answer research question (3), Alice uses the entity community view and temporal entity evolution view to understand relations among all the entities discussed in this collection.

### Identifying communities

Under the Entity Community View as in Figure 5(a), Alice identifies major communities of sub-topics around LDA topic model in this document collection: LDA models (green), LSI-related models (blue), HDP models (light-brown). Some small communities, such as those in yellow and red are discussed only a few times, thus are more marginalized in the network.

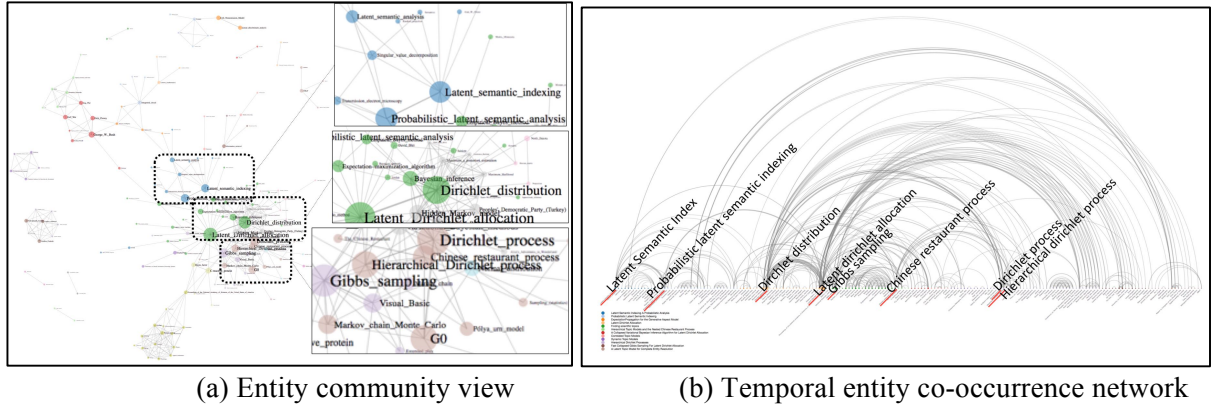

(a) Entity community view          (b) Temporal entity co-occurrence network

**Figure 5. Collective-document exploration view of topic "LDA"**

### Tracing entity evolutions

Under the Entity Evolution View, as in Figure 5(b), Alice can see how entities evolve over time. Alice pays special interests to entities with thicker arcs across papers. These entities have been frequently discussed both in its original papers and in newer papers, and this has led to a sequence of topic model names. From this sequence, Alice has a general idea of what are some older models (LSI, pLSI), what are newer models (LDA, HDP), and which new model is evolved around which old model. This provides a concise answer to question (3).

At this point, Alice successfully answers the three research questions by using the current system.

**Conclusion**

The present study proposes *LitStoryTeller*, an interactive system for visually exploring the semantic structures of scientific papers. With a screen play metaphor, the proposed system provides interactive storyline of a research article, taking advantages of a variety of techniques such as named entity recognition and comparative sentence classification. Visualizations at collection-level are also provided to aid overall reading digestion. A comprehensive case study demonstrated the usefulness of the proposed system, by answering realistic research questions from research papers. There are two limitations in the current study. First, the Microsoft API may not be able to identify very recent scientific concepts not yet included in Wikipedia. Although we have provided the function of entity customization for users, we plan to develop our own scientific entity recognizer in the future. Second, the metrics of entity importance used in present study are relatively simple, namely entity frequencies and graph centrality. In the future, we plan to propose more robust metrics such as tf*idf, to eliminate entities that are very common yet not rich in meanings in their contexts.

**Acknowledgments**

This study is supported by the NSF project "A Visual Analytic Observatory of Scientific Knowledge" (NSF 1645264).

# References

Abdul‐Rahman, A., Lein, J., Coles, K., Maguire, E., Meyer, M., Wynne, M., . . . Chen, M. (2013). *Rule‐based Visual Mappings–with a Case Study on Poetry Visualization.* Paper presented at the Computer Graphics Forum.

Alexander, E., Kohlmann, J., Valenza, R., Witmore, M., & Gleicher, M. (2014). *Serendip: Topic model-driven visual exploration of text corpora.* Paper presented at the Visual Analytics Science and Technology (VAST), 2014 IEEE Conference on.

Bikel, D. M., Miller, S., Schwartz, R., & Weischedel, R. (1997). *Nymble: a high-performance learning name-finder.* Paper presented at the Proceedings of the fifth conference on Applied natural language processing.

Blei, D. M., Ng, A. Y., & Jordan, M. I. (2003). Latent dirichlet allocation. *Journal of machine Learning research, 3*(Jan), 993-1022.

Blondel, V. D., Guillaume, J.-L., Lambiotte, R., & Lefebvre, E. (2008). Fast unfolding of communities in large networks. *Journal of statistical mechanics: theory and experiment, 2008*(10), P10008.

Borthwick, A. (1999). *A maximum entropy approach to named entity recognition.* Citeseer.

Bostock, M. (2017). https://bl.ocks.org/mbostock/4062045.

Callon, M., Courtial, J.-P., & Laville, F. (1991). Co-word analysis as a tool for describing the network of interactions between basic and technological research: The case of polymer chemsitry. *Scientometrics, 22*(1), 155-205.

Chavalarias, D., & Cointet, J.-P. (2013). Phylomemetic patterns in science evolution—the rise and fall of scientific fields. *PloS one, 8*(2), e54847.

Chen, C. (2004). Searching for intellectual turning points: Progressive knowledge domain visualization. *Proceedings of the National Academy of Sciences, 101*(suppl 1), 5303-5310.

Chen, C. (2006). CiteSpace II: Detecting and visualizing emerging trends and transient patterns in scientific literature. *Journal of the American Society for information Science and Technology, 57*(3), 359-377.

Clement, T., Plaisant, C., & Vuillemot, R. (2009). The Story of One: Humanity scholarship with visualization and text analysis. *Relation, 10*(1.43), 8485.

Correll, M., Witmore, M., & Gleicher, M. (2011). *Exploring collections of tagged text for literary scholarship.* Paper presented at the Computer Graphics Forum.

Don, A., Zheleva, E., Gregory, M., Tarkan, S., Auvil, L., Clement, T., . . . Plaisant, C. (2007). *Discovering interesting usage patterns in text collections: integrating text mining with visualization.* Paper presented at the Proceedings of the sixteenth ACM conference on Conference on information and knowledge management.

Dunne, C., Shneiderman, B., Gove, R., Klavans, J., & Dorr, B. (2012). Rapid understanding of scientific paper collections: Integrating statistics, text analytics, and visualization. *Journal of the American Society for information Science and Technology, 63*(12), 2351-2369.

Elvery, S. (2017). Narrative Charts. https://bl.ocks.org/drzax/81fff35393fb65255621fd0ab8d11bd7.

Han, J., Pei, J., Mortazavi-Asl, B., Pinto, H., Chen, Q., Dayal, U., & Hsu, M. (2001). *Prefixspan: Mining sequential patterns efficiently by prefix-projected pattern growth.* Paper presented at the proceedings of the 17th international conference on data engineering.

Jindal, N., & Liu, B. (2006a). *Identifying comparative sentences in text documents.* Paper presented at the Proceedings of the 29th annual international ACM SIGIR conference on Research and development in information retrieval.

Jindal, N., & Liu, B. (2006b). *Mining comparative sentences and relations.* Paper presented at the AAAI.

Kirschner, P. A., Buckingham-Shum, S. J., & Carr, C. S. (2012). *Visualizing argumentation: Software tools for collaborative and educational sense-making*: Springer Science & Business Media.

Kobourov, S. G. (2012). Spring embedders and force directed graph drawing algorithms. *arXiv preprint arXiv:1201.3011.*

Koch, S., John, M., Wörner, M., Müller, A., & Ertl, T. (2014). VarifocalReader—In-Depth Visual Analysis of Large Text Documents. *IEEE Transactions on Visualization and Computer Graphics, 20*(12), 1723-1732.

Liu, S., Wu, Y., Wei, E., Liu, M., & Liu, Y. (2013). Storyflow: Tracking the evolution of stories. *IEEE Transactions on Visualization and Computer Graphics, 19*(12), 2436-2445.

McCallum, A., & Li, W. (2003). *Early results for named entity recognition with conditional random fields, feature induction and web-enhanced lexicons.* Paper presented at the Proceedings of the seventh conference on Natural language learning at HLT-NAACL 2003-Volume 4.

McCurdy, N., Lein, J., Coles, K., & Meyer, M. (2016). Poemage: Visualizing the sonic topology of a poem. *IEEE Transactions on Visualization and Computer Graphics, 22*(1), 439-448.

Pantone, P. (2017). Source code for nltk.corpus.reader.comparative_sents. http://www.nltk.org/_modules/nltk/corpus/reader/comparative_sents.html.

Schneider, N., Hwa, R., Gianfortoni, P., Das, D., Heilman, M., Black, A., . . . Smith, N. A. (2010). Visualizing topical quotations over time to understand news discourse.

Tanahashi, Y., & Ma, K.-L. (2012). Design considerations for optimizing storyline visualizations. *IEEE Transactions on Visualization and Computer Graphics, 18*(12), 2679-2688.

Van Eck, N. J., & Waltman, L. (2010). Software survey: VOSviewer, a computer program for bibliometric mapping. *Scientometrics, 84*(2), 523-538.

Viegas, F. B., Wattenberg, M., & Feinberg, J. (2009). Participatory visualization with wordle. *IEEE Transactions on Visualization and Computer Graphics, 15*(6).

Vuillemot, R., Clement, T., Plaisant, C., & Kumar, A. (2009). *What's being said near "Martha"? Exploring name entities in literary text collections.* Paper presented at the Visual Analytics Science and Technology, 2009. VAST 2009. IEEE Symposium on.

Wilhelm, T., Burghardt, M., & Wolff, C. (2013). " To See or Not to See"-An Interactive Tool for the Visualization and Analysis of Shakespeare Plays.